\documentclass[conference,9pt]{IEEEtran}
\IEEEoverridecommandlockouts

\usepackage{graphicx} 
\usepackage{braket}
\usepackage{subfigure}

\usepackage{xcolor}

\usepackage{hyperref}

\usepackage{tikz}
\definecolor{custompurple}{rgb}{0.5, 0.1, 0.5}
\usepackage{circuitikz}
\usetikzlibrary{quantikz}
\usepackage{setspace} 
\usepackage{circledsteps}

\pgfkeys{/csteps/inner ysep=10pt}
\usepackage{yquant}
\useyquantlanguage{groups}

\usepackage[normalem]{ulem}

\usepackage{amsmath}
\usepackage{amssymb}
\usepackage{amsfonts}

\usepackage{algorithm}
\usepackage{algorithmicx}
\usepackage{algpseudocode}

\newtheorem{defn}{\textbf{Definition}}
\newtheorem{exam}{Example}

\newtheorem{observ}{\textbf{Observation}}

\newcommand {\F} {{\mathcal{F}}}

\newcommand {\low} {{\texttt{low}}}
\newcommand {\high} {{\texttt{high}}}
\newcommand {\idx} {{\texttt{idx}}}
\newcommand {\w} {{\texttt{wt}}}

\usepackage{authblk}

\def\BibTeX{{\rm B\kern-.05em{\sc i\kern-.025em b}\kern-.08em
    T\kern-.1667em\lower.7ex\hbox{E}\kern-.125emX}}

\begin{document}

\title{Quantum State Preparation Based on LimTDD
\thanks{Work partially supported by Innovation Program for Quantum Science and Technology under Grant No. 2024ZD0300502,  Beijing Nova Program Grant No. 20220484128 and 20240484652, and the National Natural Science Foundation of China Grant No. 12471437.}}

\author[1]{Xin Hong}
\author[1]{Chenjian Li}
\author[2]{Aochu Dai}
\author[3]{Sanjiang Li}
\author[1]{Shenggang Ying}
\author[3]{Mingsheng Ying}
\affil[1]{Key Laboratory of System Software (Chinese Academy of Sciences) and State Key Laboratory of Computer Science}
\affil[ ]{Institute of Software, Chinese Academy of Sciences, Beijing, China}
\affil[2]{Department of Computer Science and Technology, Tsinghua University, Beijing, China}
\affil[3]{Centre for Quantum Software and Information, University of Technology Sydney, Sydney, Australia}
\affil[ ]{Emails: \{hongxin, licj, yingsg\}@ios.ac.cn, dac22@mails.tsinghua.edu.cn, \{sanjiang.li, mingsheng.ying\}@uts.edu.au}
\renewcommand*{\Affilfont}{\small\it} 
\renewcommand\Authands{ and }

\maketitle

\begin{abstract}

Quantum state preparation is a fundamental task in quantum computing and quantum information processing. With the rapid advancement of quantum technologies, efficient quantum state preparation has become increasingly important. This paper proposes a novel approach for quantum state preparation based on the Local Invertible Map Tensor Decision Diagram (LimTDD). LimTDD combines the advantages of tensor networks and decision diagrams, enabling efficient representation and manipulation of quantum states. Compared with the state-of-the-art quantum state preparation method, LimTDD demonstrates substantial improvements in efficiency when dealing with complex quantum states, while also reducing the complexity of quantum circuits. Examples indicate that, in the best-case scenario, our method can achieve exponential efficiency gains over existing methods. This study not only highlights the potential of LimTDD in quantum state preparation but also provides a robust theoretical and practical foundation for the future development of quantum computing technologies.

\end{abstract}


\section{Introduction}

Quantum computing, as a cutting-edge computational technology, holds extremely significant importance. By leveraging the principles of superposition and entanglement of quantum bits (qubits), quantum computation achieves exponential acceleration in certain computational tasks, thereby demonstrating substantial advantages in solving complex problems. For example, in the field of cryptography, quantum computing can rapidly factor large integers, posing a challenge to traditional encryption methods \cite{shor1994}. In material science and drug development, quantum computing can accurately simulate the behaviour of molecules and chemical reactions, accelerating the discovery of new materials and drugs \cite{kovyrshin2025prioritizing,quantum_materials_science}. Additionally, quantum computing has a broad range of application prospects in fields such as optimisation problems and artificial intelligence \cite{Gschwendtner2025,sidford2024quantum}.

Quantum state preparation (QSP) is one of the fundamental tasks in quantum computing and quantum information processing. In quantum computing, the implementation of many quantum algorithms relies on the ability to precisely prepare specific quantum states, such as entangled states and superposition states. However, as the scale of quantum systems increases, the complexity of quantum states grows exponentially, making the efficient preparation of quantum states a highly challenging problem.

In recent years, with the rapid development of quantum technologies, significant progress has been made in developing (quantum) algorithms for QSP. Many methods have been established to study the state preparation of special quantum states, such as sparse quantum states \cite{gleinig2021efficient,mao2024toward, PhysRevA.110.032609}. Methods based on the gate decomposition \cite{plesch_quantum-state_2011, shende_synthesis_2006,Iten_Colbeck_Kukuljan_Home_Christandl_2016} and uniformly controlled rotations \cite{mottonen2004transformation} have been proposed for the preparation of general quantum states. Some works focus on preparing a quantum state with the minimal number of ancilla qubits \cite{bergholm_quantum_2005}, and some works focus on preparing a quantum state with the minimal circuit depth \cite{zhang_quantum_2022}. At the same time, theoretical upper and lower bounds have also been identified \cite{sun2023asymptotically} for QSP. However, most of the methods are established on explicit representations of the quantum state, such as the vector representation, whose size grows exponentially as the number of qubits increases, restricting the size of the quantum state that can be prepared. 

Decision diagrams, as an efficient mathematical tool and a compact data structure, have been widely used in classical circuit synthesis \cite{5227151} and verification \cite{9707932}. In recent years, researchers have begun to explore the application of decision diagrams in the quantum field. By adapting classical decision diagrams to support quantum operations or developing new decision diagrams, researchers have made progress in efficient simulation and verification of quantum circuits \cite{burgholzer2021qcec, ftdd10528748}. There are also several works using decision diagrams to conduct QSP, such as \cite{mozafari_automatic_2020, mozafari_efficient_2022,tanaka_quantum_2024}. Due to the compactness of decision diagrams, these algorithms demonstrate high efficiency and are capable of supporting the preparation of relatively large-scale quantum states. Mozafari et al. proposed an efficient algorithm for the preparation of uniform quantum states \cite{mozafari_automatic_2020}. The algorithm in \cite{mozafari_efficient_2022} uses one auxiliary qubit, with complexity related to the number of paths in the decision diagram, while the algorithm in \cite{tanaka_quantum_2024} employs multiple auxiliary qubits, with complexity related to the number of nodes in the decision diagram. Although these algorithms are highly efficient, the compression efficiency of the decision diagrams they employ may potentially affect the effectiveness of the final state preparation. 


Recently, a new decision diagram, Local Invertible Map Tensor Decision Diagram (LimTDD) \cite{hong_limtdd_2025}, has been proposed and shows great compactness compared to other decision diagrams. LimTDD combines the strengths of TDD and LIMDD, can be used to represent any tensors and can identify the isomorphic structure between tensors, achieving an exponential gap between other decision diagrams, such as TDD and LIMDD, in the best-case scenario.

This paper introduces a novel quantum state preparation algorithm based on LimTDD. By leveraging the extreme compression efficiency of LimTDD, our algorithm can handle large-scale quantum states more efficiently, while also reducing the number of quantum gates and circuit depth. This work not only highlights the potential of LimTDD in QSP but also provides a robust foundation for the future development of quantum computing technologies. We have developed an interface that converts state vectors into LimTDDs and then synthesises these representations into executable state preparation circuits. This capability is designed to streamline the workflow for quantum physicists and experimentalists, enabling them to efficiently generate the quantum states required for their research. In future work, we plan to integrate our algorithm into widely used quantum computing frameworks, such as Qiskit, to further accelerate and standardise the QSP process.



The structure of this paper is as follows. In section~\ref{sec:background}, we provide basic concepts of quantum computing and QSP. In section~\ref{sec:LimTDD}, we introduce how to represent quantum states as LimTDDs.  In section~\ref{sec:Algorithms}, we give the basic algorithms for QSP using LimTDD. A detailed example is given in section \ref{sec:exp}. The complexity of the algorithm is analysed in section \ref{sec:complexity}. Then, we will conduct experiments to carefully analyse the performance of our algorithm in section~\ref{sec:experiments}. At last, in section~\ref{sec:conclusion}, we conclude the paper.

\section{Background}\label{sec:background}

In this section, we give basic background on quantum computing and quantum state preparation.

\subsection{Quantum Computing}

\subsubsection{Quantum States}
Quantum computing is a paradigm that leverages the principles of quantum mechanics to perform computations. Unlike classical computing, which operates on classical bits that can be in a state of either $0$ or $1$, quantum computing utilises quantum bits (qubits) that can exist in superpositions of states. A qubit is described by a two-dimensional complex vector space, with the computational basis states denoted as $\ket{0}$ and $\ket{1}$. These basis states are orthonormal, satisfying the conditions:

\[
\langle 0 | 0 \rangle = \langle 1 | 1 \rangle = 1 \quad \text{and} \quad \langle 0 | 1 \rangle = \langle 1 | 0 \rangle = 0.
\]

A general state of a qubit can be represented as a linear combination of these basis states:

\[
\ket{\psi} = \alpha \ket{0} + \beta \ket{1}.
\]

where $\alpha$ and $\beta$ are complex numbers known as probability amplitudes, satisfying the normalisation condition $|\alpha|^2 + |\beta|^2 = 1$. This superposition property allows a qubit to represent multiple states simultaneously, providing a fundamental advantage over classical bits.

For multi-qubit systems, the state space grows exponentially as the number of qubits increases. For instance, an $n$-qubit system is described by a $2^n$-dimensional complex vector, with the computational basis states denoted as $\ket{k}$, where $k$ is a binary string of length $n$. For example, the state $\ket{000}$ represents the state where all three qubits are in the $\ket{0}$ state, and can be written as:

\[
\ket{000} = \ket{0} \otimes \ket{0} \otimes \ket{0}.
\]

\begin{exam}\label{exp:q_state}
Consider the 3-qubit quantum state $\frac{1}{\sqrt{6}}(\ket{000}+\ket{001}+\frac{1}{\sqrt{2}}\ket{010}-\frac{1}{\sqrt{2}}\ket{011}-\ket{100}-\frac{1}{\sqrt{2}}\ket{101}+\frac{1}{\sqrt{2}}\ket{110}+\ket{111})$. This state can be represented as an 8-dimensional vector: $\frac{1}{\sqrt{6}}[1,1,\frac{1}{\sqrt{2}},-\frac{1}{\sqrt{2}},-1,-\frac{1}{\sqrt{2}},\frac{1}{\sqrt{2}},1]^T$. Here, each component of the vector corresponds to a specific computational basis state.
\end{exam}

\subsubsection{Quantum Gates}

Quantum computations are performed through the application of quantum gates, which are unitary transformations on the qubits. Common quantum gates include:

\begin{itemize}
    \item \textbf{Hadamard gate ($H$ gate)}: This gate creates a superposition state by transforming the basis states $\ket{0}$ and $\ket{1}$ into equal superpositions:
    \[
    H\ket{0} = \frac{1}{\sqrt{2}}(\ket{0} + \ket{1}) \quad \text{and} \quad H\ket{1} = \frac{1}{\sqrt{2}}(\ket{0} - \ket{1}).
    \]

    
    \item \textbf{Pauli-$Z$ gate ($Z$ gate)}: This gate introduces a phase flip to the $\ket{1}$ state and remain the state $\ket{0}$ unchanged:
    \[
    Z\ket{0} =\ket{0} \quad \text{and} \quad Z\ket{1}=-\ket{1}.
    \]

    \item \textbf{Controlled-$X$ gate ($CX$ gate)}: Also known as the CNOT gate, this gate performs a NOT operation on a target qubit conditioned on the state of a control qubit. For example, if the control qubit is in the state $\ket{1}$, the target qubit's state is flipped:
    \[
    CX(\ket{0} \otimes \ket{\psi}) \!=\! \ket{0} \otimes \ket{\psi}\  \text{and}\ CX(\ket{1} \otimes \ket{\psi}) \!=\! \ket{1} \otimes X\ket{\psi}.
    \]
\end{itemize}

\subsubsection{Quantum Circuits}

Quantum circuits are constructed by combining these basic gates in specific sequences to implement various quantum algorithms. Each gate is applied to one or more qubits, and the sequence of gates determines the overall transformation applied to the qubits. The output of a quantum circuit is typically measured in the computational basis, providing the result of the computation.

Quantum circuits can be represented graphically, with qubits as horizontal lines and gates as symbols applied to these lines. This graphical representation helps visualise the sequence of operations and the flow of information through the circuit.

\begin{figure}[htbp]
\centering
\begin{tikzpicture}
  \begin{yquant}[register/minimum height=8mm, operator/separation=5mm, control style={radius=2pt}, subcircuit box style={dashed}]
    qubit {$\ket{q_2}$} q2;
    qubit {$\ket{q_1}$} q1;
    qubit {$\ket{q_0}$} q0;
    box {$H$} q2;
    box {$H$} q1;
    box {$H$} q0;
    z q0 | q1;
    z q0 | q2;
    z q1 | q2;
    
  \end{yquant}
\end{tikzpicture}
\caption{Example of a quantum circuit with Hadamard gates and $CZ$ gates.}
    \label{fig:quantum_circuit}
\end{figure}
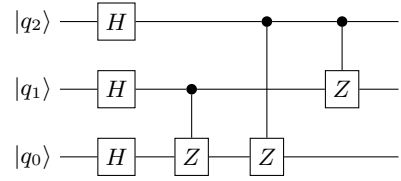

Fig. \ref{fig:quantum_circuit} gives an example of quantum circuits. In this example, the quantum circuit consists of Hadamard gates and $CZ$ gates applied to three qubits. The Hadamard gates create superposition states, and the $CZ$ gates introduce entanglement. The specific sequence of gates determines the transformation applied to the qubits, which can be used to implement various quantum algorithms.

\subsection{Quantum State Preparation}

Quantum state preparation is a fundamental task in quantum computing and quantum information processing, aiming to transform a given initial state into a desired target quantum state. This process is crucial for the implementation of various quantum algorithms, as many algorithms require specific quantum states as inputs to achieve their computational advantages.

\subsubsection{Definition and Objective}

Formally, quantum state preparation can be defined as follows. Given an initial state, typically $\ket{0}^{\otimes n}$, and a target quantum state $\ket{\psi_v} = \sum_{k=0}^{2^n-1} v_k \ket{k}$, where $v = (v_0, v_1, \dots, v_{2^n-1})^T \in \mathbb{C}^{2^n}$ is a normalized vector ($\|v\|_2 = 1$) representing the amplitudes of the target state in the computational basis, the objective of QSP is to construct a quantum circuit that transforms the initial state into the target state $\ket{\psi_v}$.

\subsubsection{Challenges and Importance}

As the number of qubits $n$ increases, the complexity of representing and manipulating quantum states grows exponentially. Specifically, a general $n$-qubit state requires $2^n$ complex amplitudes to be fully described, making the explicit representation and preparation of arbitrary quantum states computationally intractable in general cases. Efficient QSP is thus a highly challenging problem, especially for large quantum systems. However, it is also essential for practical quantum computing applications, as the efficiency of QSP directly impacts the feasibility and performance of many quantum algorithms.

\subsubsection{Existing Approaches and Limitations}

In \cite{mozafari_efficient_2022}, Mozafari et al. proposed an efficient algorithm for QSP using decision diagrams. The proposed algorithm starts from the $\ket{0}$ state and prepares the quantum state by traversing the decision diagram path by path. It uses an ancilla qubit to mark the paths that have already been processed. By using this auxiliary qubit as a control qubit, the subsequent preparation process will not affect the paths that have already been prepared. As a result, its time complexity and the gate complexity of the resulting circuit are proportional to the number of paths in the decision diagram. 

However, the decision diagram used in \cite{mozafari_efficient_2022} is a multi-terminal Algebraic Decision Diagram (ADD), which has a relatively low compactness representing quantum states. For example, the quantum state given in Example \ref{exp:q_state} requires seven paths to be represented as an ADD as shown in Fig. \ref{fig:limtdd} (a), while for other decision diagrams such as LimTDD, only two (reduced) paths are needed.

The algorithm presented in this paper is inspired by the work of \cite{mozafari_efficient_2022}. But differs significantly in details, especially that we use an ancilla qubit to indicate the open or closed state of a node rather than a path. This difference is determined by the inherent characteristics of the two different decision diagrams.

\section{Representing Quantum State Using LimTDD}\label{sec:LimTDD}

LimTDD is a highly compact decision diagram designed for representing and operating tensors and tensor networks in an efficient way. In this paper, we use the data structure LimTDD to represent quantum states, which is capable of representing a quantum state with a very high compression ratio. This feature ensures efficient preparation and the preparation of large-scale quantum states.

\subsection{LimTDD}

The efficient representation of LimTDD is established on the isomorphism of quantum states.

\vspace{0.2em}
\begin{defn}[LIM, Quantum State Isomorphism]
An $n$-qubit Local Invertible Map (LIM) is an operator $O$ of the form 
\begin{equation}\label{eq:limO}
    O = \lambda O_n \otimes \cdots \otimes O_1, 
\end{equation} 
where $\lambda\in \mathbb C$ is a complex number and each $O_i$ is an invertible $2\times 2$ matrix. The set of all such maps is denoted as $\mathcal{M}(n)$, and the set of all LIMs, which is a group, is defined as 
\begin{equation}\label{eq:limCalO}
\mathcal{M} = \bigcup_{n\in N}{\mathcal{M}(n)}.
\end{equation} 
Two $n$-qubit quantum states $\ket{\Psi}$ and $\ket{\Phi}$ are said to be \emph{isomorphic} if $\ket{\Phi} = O \ket{\Psi}$ for some $O \in \mathcal{M}(n)$.
\end{defn}

\vspace{0.2em}
\begin{defn}[LimTDD]\label{def:limtdd}
	Let $\mathcal{G}$ be a subgroup of $\mathcal{M}$. A $\mathcal{G}$-LimTDD $\mathcal{F}$ over a set of indices $S$ is a rooted, weighted, and directed acyclic graph 
	$\mathcal{F} = (V, E, \idx, \low, \high, \w)$ defined as follows:
	\begin{itemize}
		\item $V$ is a finite set of nodes which consists of non-terminal nodes $V_{NT}$ and a terminal node $v_T$ labelled with integer 1. Denote by $r_\mathcal{F}$ the unique root node of $\mathcal{F}$;
		\item $\idx: V_{NT} \rightarrow S$ assigns each non-terminal node an index in $S$. We call $\idx(r_\mathcal{F})$ the top index of $\F$, if $r_\mathcal{F}$ is not the terminal node;
		\item both $\low$ and $\high$ are mappings in $V_{NT} \rightarrow V$, which map each non-terminal node to its 0- and 1-successors, respectively;
		\item $E = \{(v, \low(v)), (v, \high(v)) : v\in V_{NT}\}$ is the set of edges, where $(v, \low(v))$ and $(v, \high(v)) $ are called the low- and high-edges of $v$, respectively. For simplicity, we also assume the root node $r_\mathcal{F}$ has a unique incoming edge, denoted  $e_r$, which has no source node;
		\item $\w: E\rightarrow \mathcal{G}$ assigns each edge a weight in $\mathcal{G}$.  $\w(e_r)$ is called the weight of $\mathcal{F}$, and denoted $w_\mathcal{F}$.  
	\end{itemize} 

\end{defn}

When representing a quantum state, the semantics of the terminal node is defined to be $\ket{v_T} = 1$, the semantics of an edge $e$, directing to a node $v$, is defined to be 
$$
\ket{e} = \w(e) \cdot \ket{v},
$$
and the semantics of a non-terminal node $v$ is defined to be 
$$
\ket{v} = \ket{0}\otimes \big|(v,\low(v))\big\rangle + \ket{1}\otimes \big|(v,\high(v))\big\rangle\\
$$

Note that, when representing quantum states, every index corresponds to a qubit. In this paper, for convenience, we always assume that the top index corresponds to the most significant qubit (that is, $q_{n-1}$, for an $n$-qubit quantum state), and index of the bottom non-terminal node corresponds to the least significant qubit (that is, $q_0$), with the other qubits arranged in sequence. Sometimes, to identify the qubits, we use notations such as $\ket{0}_{n-1}$ and $\ket{0}_0$. The subgroup $\mathcal{G}$ is set to be XP-Operators in \cite{hong_limtdd_2025}, however, the algorithm in this paper works for any subgroup of $\mathcal{M}$.

\vspace{-1em}

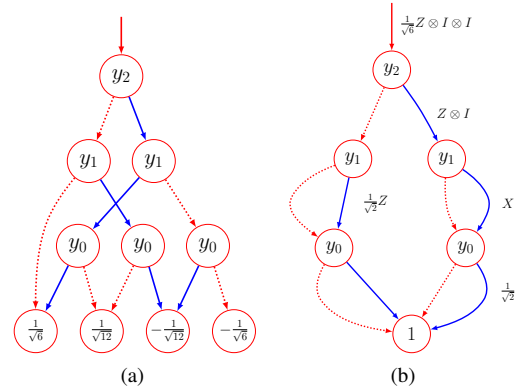
\begin{figure}[htbp]
    \centering
    \subfigure[]{
    \resizebox{0.18\textwidth}{!}{




\begin{tikzpicture}[
>=latex,
line join=bevel,
every node/.style={minimum size=1.4cm, inner sep=0pt,thick, font=\Huge}, 
every path/.style={line width=1.5pt} 
]


\node (1) at (21.667bp,21.667bp) [draw=red,circle,font=\LARGE] {$\frac{1}{\sqrt{6}}$};
  \node (-1) at (204.67bp,21.667bp) [draw=red,circle,font=\LARGE] {$-\frac{1}{\sqrt{6}}$};
  \node (2) at (82.667bp,21.667bp) [draw=red,circle,font=\LARGE] {$\frac{1}{\sqrt{12}}$};
  \node (-2) at (143.67bp,21.667bp) [draw=red,circle,font=\LARGE] {$-\frac{1}{\sqrt{12}}$};
  \node (y01) at (120.67bp,100.49bp) [draw=red,circle] {$y_0$};
  \node (y02) at (180.67bp,100.49bp) [draw=red,circle] {$y_0$};
  \node (y03) at (60.667bp,100.49bp) [draw=red,circle] {$y_0$};
  \node (y10) at (70.667bp,178.8bp) [draw=red,circle] {$y_1$};
  \node (y11) at (130.67bp,178.8bp) [draw=red,circle] {$y_1$};
  \node (y2) at (100.67bp,257.11bp) [draw=red,circle] {$y_2$};
  \node (-0) at (100.67bp,332.26bp) [draw,draw=none] {};
  \draw [red,->,dotted] (y01) ..controls (107.14bp,72.152bp) and (101.89bp,61.528bp)  .. (2);
  \draw [blue,->] (y01) ..controls (128.94bp,71.853bp) and (131.7bp,62.632bp)  .. (-2);
  \draw [red,->,dotted] (y02) ..controls (189.22bp,72.102bp) and (192.18bp,62.628bp)  .. (-1);
  \draw [blue,->] (y02) ..controls (167.5bp,72.152bp) and (162.38bp,61.528bp)  .. (-2);
  \draw [blue,->] (y03) ..controls (46.958bp,72.485bp) and (41.488bp,61.71bp)  .. (1);
  \draw [red,->,dotted] (y03) ..controls (68.581bp,71.853bp) and (71.221bp,62.632bp)  .. (2);
  \draw [red,->,dotted] (y10) ..controls (47.918bp,151.69bp) and (37.26bp,136.79bp)  .. (31.667bp,121.64bp) .. controls (23.786bp,100.31bp) and (21.367bp,74.739bp)  .. (1);
  \draw [blue,->] (y10) ..controls (88.277bp,150.92bp) and (96.192bp,138.84bp)  .. (y01);
  \draw [red,->,dotted] (y11) ..controls (148.28bp,150.92bp) and (156.19bp,138.84bp)  .. (y02);
  \draw [blue,->] (y11) ..controls (107.03bp,152.03bp) and (93.66bp,137.46bp)  .. (y03);
  \draw [red,->,dotted] (y2) ..controls (89.863bp,228.63bp) and (86.014bp,218.84bp)  .. (y10);
  \draw [blue,->] (y2) ..controls (111.47bp,228.63bp) and (115.32bp,218.84bp)  .. (y11);
  \draw [red,->] (-0) ..controls (100.67bp,307.06bp) and (100.67bp,298.38bp)  .. (y2);
\end{tikzpicture}

%
%
}
    }
    \subfigure[]{
    \resizebox{0.18\textwidth}{!}{




\begin{tikzpicture}[
>=latex,
line join=bevel,
every node/.style={minimum size=1.4cm, inner sep=0pt,thick, font=\LARGE}, 
every path/.style={line width=1.5pt} 
]
\node (1) at (128.67bp,18.0bp) [draw=red,circle,font=\Huge] {$1$};
  \node (y01) at (46.668bp,108.9bp) [draw=red,circle,font=\Huge] {$y_0$};
  \node (y02) at (185.67bp,108.9bp) [draw=red,circle,font=\Huge] {$y_0$};
  \node (y10) at (66.668bp,202.96bp) [draw=red,circle,font=\Huge] {$y_1$};
  \node (y11) at (165.67bp,202.96bp) [draw=red,circle,font=\Huge] {$y_1$};
  \node (y2) at (107.67bp,297.02bp) [draw=red,circle,font=\Huge] {$y_2$};
  \node (-0) at (107.67bp,387.93bp) [draw,draw=none] {};
  \draw [red,->,dotted] (y01) ..controls (29.551bp,79.314bp) and (25.128bp,64.774bp)  .. (32.668bp,54.0bp) .. controls (47.583bp,32.686bp) and (76.911bp,24.306bp)  .. (1);
  \definecolor{strokecol}{rgb}{0.0,0.0,0.0};
  \pgfsetstrokecolor{strokecol}
  \draw (53.668bp,61.875bp) node {};
  \draw [blue,->] (y01) ..controls (73.917bp,78.36bp) and (94.182bp,56.39bp)  .. (1);
  \draw (116.67bp,61.875bp) node {};
  \draw [red,->,dotted] (y02) ..controls (166.06bp,77.316bp) and (153.94bp,58.425bp)  .. (1);
  \draw (180.67bp,61.875bp) node {};
  \draw [blue,->] (y02) ..controls (205.77bp,80.723bp) and (212.16bp,65.699bp)  .. (204.67bp,54.0bp) .. controls (194.36bp,37.899bp) and (174.56bp,29.102bp)  .. (1);
  \draw (230.29bp,61.875bp) node {$\frac{1}{\sqrt{2}}$};
  \draw [red,->,dotted] (y10) ..controls (31.234bp,190.19bp) and (11.932bp,180.12bp)  .. (2.6676bp,163.81bp) .. controls (-5.0407bp,150.24bp) and (6.2358bp,136.64bp)  .. (y01);
  \draw (23.668bp,155.93bp) node {};
  \draw [blue,->] (y10) ..controls (59.707bp,169.93bp) and (56.313bp,154.3bp)  .. (y01);
  \draw (90.543bp,155.93bp) node {$\frac{1}{\sqrt{2}} Z$};
  \draw [red,->,dotted] (y11) ..controls (163.61bp,171.29bp) and (164.02bp,158.88bp)  .. (166.67bp,148.06bp) .. controls (167.5bp,144.65bp) and (168.68bp,141.19bp)  .. (y02);
  \draw (187.67bp,155.93bp) node {};
  \draw [blue,->] (y11) ..controls (193.01bp,184.7bp) and (203.51bp,175.28bp)  .. (208.67bp,163.81bp) .. controls (212.85bp,154.5bp) and (210.23bp,144.22bp)  .. (y02);
  \draw (230.42bp,155.93bp) node {$X$};
  \draw [red,->,dotted] (y2) ..controls (93.74bp,264.75bp) and (86.177bp,247.77bp)  .. (y10);
  \draw (110.67bp,249.99bp) node {};
  \draw [blue,->] (y2) ..controls (125.36bp,273.0bp) and (130.96bp,265.2bp)  .. (135.67bp,257.87bp) .. controls (140.87bp,249.76bp) and (146.13bp,240.69bp)  .. (y11);
  \draw (172.92bp,249.99bp) node {$Z\otimes I$};
  \draw [red,->] (-0) ..controls (107.67bp,358.85bp) and (107.67bp,343.46bp)  .. (y2);
  \draw (154.54bp,344.05bp) node {$\frac{1}{\sqrt{6}} Z\otimes I\otimes I$};
\end{tikzpicture}

%
%
}
    }
    \caption{An example of Multiple-terminal ADD \cite{mozafari_efficient_2022} and LimTDD representing the quantum state $\frac{1}{\sqrt{6}}[1,1,\frac{1}{\sqrt{2}},-\frac{1}{\sqrt{2}},-1,-\frac{1}{\sqrt{2}},\frac{1}{\sqrt{2}},1]^T$. We omit the weight $1$ and $1\cdot I^{\otimes k}$ in the figure of LimTDD. 
    }
    \label{fig:limtdd}
\end{figure}

\begin{exam}\label{exp:limtdd}
Fig. \ref{fig:limtdd} (b) gives an example of LimTDD, which represents the quantum state shown in example \ref{exp:q_state}. In this figure, we use dotted red lines to represent the low edges and solid blue lines to represent the high edges. We refer to the node with index $y_2$ as the $y_2$ node. The node on the left with index $y_1$ (or $y_0$) is called the $y_1$ (or $y_0$) node. The node on the right with index $y_1$ (or $y_0$) is called the $y_1'$ (or $y_0'$) node. The indices $y_2,\ y_1$ and $y_0$ corresponds to the qubits $q_2,\ q_1$ and $q_0$. Then, ignoring the normalisation coefficients,
\begin{itemize}
    \item $y_0$ node represents the quantum state $\ket{y_0} = \ket{0}+\ket{1}$, $y_1$ node represents the quantum state $\ket{y_1} = \ket{0}\ket{y_0}+\frac{1}{\sqrt{2}}\ket{1}(Z \ket{y_0}) = \ket{00}+\ket{01}+\frac{1}{\sqrt{2}}\ket{10}-\frac{1}{\sqrt{2}}\ket{11}$.\vspace{0.2em}
    \item Similarly, the $y_0'$ node represents the quantum state $\ket{y_0'} = \ket{0}+\frac{1}{\sqrt{2}}\ket{1}$, $y_1'$ node represents the quantum state $\ket{y_1'} = \ket{0}\ket{y_0'}+\ket{1}(X \ket{y_0'}) = \ket{00}+\frac{1}{\sqrt{2}}\ket{01}+\frac{1}{\sqrt{2}}\ket{10}+\ket{11}$.\vspace{0.2em}
    \item Then the $y_2$ node represent the quantum state $\ket{y_2} = \ket{0}\ket{y_1} + \ket{1}(Z\otimes I \ket{y_1'}) = \ket{000}+\ket{001}+\frac{1}{\sqrt{2}}\ket{010}-\frac{1}{\sqrt{2}}\ket{011}+\ket{100}+\frac{1}{\sqrt{2}}\ket{101}-\frac{1}{\sqrt{2}}\ket{110}-\ket{111}$.\vspace{0.2em}
    \item Finally, the entire LimTDD represents the quantum state $\frac{1}{\sqrt{6}} Z\otimes I\otimes I \ket{y_2} = \frac{1}{\sqrt{6}}(\ket{000}+\ket{001}+\frac{1}{\sqrt{2}}\ket{010}-\frac{1}{\sqrt{2}}\ket{011}-\ket{100}-\frac{1}{\sqrt{2}}\ket{101}+\frac{1}{\sqrt{2}}\ket{110}+\ket{111})$.\vspace{0.2em}
\end{itemize}

\end{exam}


\begin{defn}[Reduced paths]\label{def:red_path}
Let $\mathcal{F}$ be a decision diagram. The \emph{reduced diagram} of $\mathcal{F}$ is obtained by merging the edges between any two nodes in $\mathcal{F}$. We also call paths within this reduced diagram \emph{reduced paths} of $\mathcal{F}$.
\end{defn}

In the example provided, the $y_1$ node has identical 0-successor and 1-successor nodes, meaning its two outgoing edges are merged when determining the reduced paths. The same applies to the $y_1', y_0$, and $y_0'$ nodes. Consequently, the LimTDD features only 2 reduced paths, whereas the multiple-terminal ADD comprises 7 reduced paths.



\vspace{0.2em}
In this paper, we employ a specific graphical representation for LimTDDs. A node in a LimTDD is uniquely determined by its index, its two successors, and the weights on the edges connecting to those successors. We denote this as:
\[\Circled{v_{0}}\overset{w_0}{\dashleftarrow}\Circled{v}\xrightarrow[]{w_1}\Circled{v_{1}}.\] 
Similarly, a LimTDD is uniquely determined by its root node and the weight on the incoming edge. This can be represented as:
\[\Big(w_\F,\ \Circled{v_{0}}\overset{w_0}{\dashleftarrow}\Circled{v}\xrightarrow[]{w_1}\Circled{v_{1}}\Big),\] or more succinctly as: \[\xrightarrow[]{w_\F}\Circled{v}.\]

\section{LimTDD Based Quantum State Preparation} \label{sec:Algorithms}

In this section, we introduce our method for Quantum State Preparation. Our method is mainly based on the following three observations.

\subsection{Basic constructions}


\begin{observ}
    Let $\F = \xrightarrow[]{\lambda O_n\otimes \cdots \otimes O_1}\Circled{v} $ be a LimTDD representing the quantum state $\ket{\psi}$. Then $\ket{\psi} = \lambda O_n\otimes \cdots \otimes O_1 \ket{v}$. Applying the operator $(O_n\otimes \cdots \otimes O_1)^\dag$ to $\ket{\psi}$ can reduce it to $\lambda \cdot \ket{v}$, which can be represented using a LimTDD with the root node $v$ and incoming edge weight $\lambda$, that is $\xrightarrow[]{\lambda }\Circled{v}$. In other words, applying a $(O_n\otimes \cdots \otimes O_1)^\dag$ to the LimTDD (i.e., contracting each index $y_i$ with an operator $O_i$) can cancel the operator on the incoming edge of the LimTDD.
\end{observ}\vspace{0.2em}




\begin{observ}
Let $\F = \Big(w,\ \Circled{v_{0}}\overset{I}{\dashleftarrow}\Circled{v}\xrightarrow[]{\lambda O_n\otimes \cdots \otimes O_1}\Circled{v_{1}}\Big)$ be a LimTDD representing the quantum state $\ket{\psi}$. Suppose the weight of $\F$ is a complex number $w$. Then $\ket{\psi} = w \cdot (\ket{0}\ket{v_0} + \ket{1}\otimes (\lambda O_n\otimes \cdots \otimes O_1 \ket{v_1}))$. Applying the operator $(O_n\otimes \cdots \otimes O_1)^\dag$ controlled by the qubit corresponding to the root node of $\F$ to $\ket{\psi}$ can reduce it to $w\cdot (\ket{0}\ket{v_0} + \lambda \ket{1}\ket{v_1})$, which can be represented using the LimTDD $\Big(w,\ \Circled{v_{0}}\overset{I}{\dashleftarrow}\Circled{v}\xrightarrow[]{\lambda}\Circled{v_{1}}\Big)$. In other words, the operator on the high-edge of the node has been cancelled.
\end{observ}
\vspace{0.2em}


\begin{observ}
Let $\F = \Big(w,\ \Circled{v_{0}}\overset{w_0}{\dashleftarrow}\Circled{v}\xrightarrow[]{w_1}\Circled{v_{0}}\Big)$ be a LimTDD representing the quantum state $\ket{\psi}$. Suppose $w$, $w_0$, $w_1$ are complex numbers and $w_0 \neq 0$. Then $\ket{\psi} = w \cdot (w_0\ket{0}+w_1\ket{1})\ket{v_0}$. Denote $c = w_1/w_0$. Applying the unitary operator  
$\frac{1}{\sqrt{1+|c|^2}}\left[\begin{array}{cccc} 
		1 & c^\dag\\ 
		-c & 1\\
\end{array}\right]$
to $\ket{\psi}$ can reduce it to $w\cdot w_0 \cdot \sqrt{1+|c|^2} \cdot \ket{0}\ket{v_0}$, which can be represented using the LimTDD $\Big(w\cdot w_0 \cdot \sqrt{1+|c|^2},\ \Circled{v_{0}}\overset{1}{\dashleftarrow}\Circled{v}\xrightarrow[]{0}\Circled{v_{0}}\Big)$. In other words, the complex weights on the two outgoing edges of $v$ have been reduced to $[1,0]$, corresponding to $\ket{0}$.
\end{observ}

\subsection{Branch processing}

The simplifications introduced above apply primarily to the root node. In this subsection, we extend these techniques to non-terminal nodes by using an ancilla qubit to selectively ``close'' parts of the decision diagram, thereby exposing and processing other nodes as if they were root nodes.

\begin{defn}[Branch Condition]
A non-terminal node is termed a branch node if its 0-successor and 1-successor are distinct. The branch condition for a node along a path is defined by the values of all branch nodes preceding it on that path.
\end{defn}

Consider the LimTDD given in Fig. \ref{fig:limtdd} (b). Here, node $y_2$ is a branch node. The branch condition for the $y_1$ node (the left one) is $\ket{0}_2$, while for the $y_1'$ node (the right one), it is $\ket{1}_2$.


\begin{defn}[Open/Closed Node]\label{def:oc_node}
Let $\mathcal{F}$ be a LimTDD, representing the quantum state $\ket{\psi_v} = \sum_{k=0}^{2^n-1} v_k \ket{k}$. Introducing an ancilla qubit $q_a$ and marking each computational basis state $\ket{k}$ with $\ket{b_k}_a$ gives the state $\sum_{k=0}^{2^n-1} v_k \ket{b_k}_a\ket{k}$, where $b_k \in \{0,1\}$. A path, corresponding to the computational basis state $\ket{k}$, is called open (closed, resp.) by $q_a$, if $b_k = 1$ ($0$, resp.). 
A node $v$ on a path is called open (closed, resp.) if all paths that consist of the prefix path up to this node, as well as any suffix path rooted at this node, are open (closed, resp.), i.e., marked as $\ket{1}_a$ ($\ket{0}_a$, resp.). A LimTDD is called open (closed, resp.) if its root node is open (closed, resp.).

\end{defn}


\vspace{0.2em}
With these definitions, we can now use the ancilla qubit to manipulate the LimTDD.
Suppose a node $v$ and its successors are the only open nodes. We use $q_a$ as a control qubit to perform operations on $v$ without affecting other parts of the decision diagram.

For instance, consider the LimTDD $\F = \Big(w,\ \Circled{v_{0}}\overset{I}{\dashleftarrow}\Circled{v}\xrightarrow[]{w_1}\Circled{v_{1}}\Big)$, representing the quantum state $\ket{\psi} = w \cdot (\ket{0}_b\ket{v_0} + \lambda \ket{1}_b\ket{v_1})$. Let $q_b$ and $q_c$ be the qubits corresponding to nodes $v$ and $v_0$, respectively. By setting the ancilla qubit $q_a$ to $\ket{1}_a$, the entire LimTDD is initially open. Applying a CX gate controlled by $\ket{1}_b$ yields the state $w \cdot (\ket{1}_a\ket{0}_b\ket{v_0} + \lambda \ket{0}_a\ket{1}_b\ket{v_1})$, effectively closing node $v_1$ and opening node $v_0$.

We can then use $\ket{1}_a$ to control further operations on $v_0$. Suppose $\Circled{v_{00}}\overset{I}{\dashleftarrow}\Circled{v_0}\xrightarrow[]{\lambda O_n\otimes \cdots \otimes O_1}\Circled{v_{01}}$, such that $\ket{v_0}=\ket{0}_c\ket{v_{00}}+\lambda\ket{1}_c(O_n\otimes \cdots \otimes O_1\ket{v_{01}})$. Applying an $(O_n\otimes \cdots \otimes O_1)^\dag$ gate controlled by $\ket{1}_a\ket{1}_{c}$ transforms the state to $w \cdot (\ket{1}_a\ket{0}\ket{v_0^\prime} + \lambda \ket{0}_a\ket{1}\ket{v_1})$, where $\ket{v_0^\prime}=\ket{0}\ket{v_{00}}+\lambda\ket{1}\ket{v_{01}}$, depicted as $\Circled{v_{00}}\overset{I}{\dashleftarrow}\Circled{v_0^\prime}\xrightarrow[]{\lambda}\Circled{v_{01}}$. \vspace{0.4em}

After processing the $v_0$ branch, applying an $X$ gate to $q_a$ switches the state to $w \cdot (\ket{0}_a\ket{0}_b\ket{v_0^\prime} + \lambda \ket{1}_a\ket{1}_b\ket{v_1})$, closing node $v_0$ and opening node $v_1$ for subsequent processing. To ensure the ancilla qubit $q_a$ is restored to its initial state, we apply a $CX$ gate with the control condition $\ket{0}_b$. This operation changes the state to $w \cdot \ket{1}_a(\ket{0}_b\ket{v_0^\prime} + \lambda \ket{1}_b\ket{v_1'})$, thus restoring $q_a$ to $\ket{1}_a$. It should be noted that if the nodes $v_0$, $v_1$, and their subsequent nodes have all been fully processed, the state should have the form $w \cdot \ket{1}_a(w_0\ket{0}_b + w_1 \ket{1}_b)\ket{0}^{\otimes k}$, for some $k$. At this time, Observation 3 can be applied to reduce the weights $w_0$ and $w_1$ and return the qubit $q_b$ to $\ket{0}$.

\subsection{Algorithm}
\label{sec:algorithm}

In our algorithm (pseudocode described in Algorithm \ref{alg:State_Pre}), we first cancel the operator on the incoming edge of the root node. Then, we enter a recursive process. For every node, we first cancel the operator on its high-edge, then process its 0-successor, followed by its 1-successor, and finally adjust the weights on the outgoing edges of the current node.

During this process, we use an auxiliary qubit to mark the current node $v$ being processed, i.e., the open part of the decision diagram. Whenever we encounter a branch node, we first use a multi-controlled $X$ gate to close its 1-successor. The control condition should be set to the branch condition of $\high(v)$. Then we process the 0-successor. After that, we flip the open/close condition of the two successors with the branch condition of $v$ and process the 1-successor. Finally, we open the 0-successor using the branch condition of $\low(v)$, making the original node open. Then we backtrack upwards. Note that when we finish the process, the root node will be open, meaning that the ancilla qubit has been returned to the $\ket{1}_a$ state. The circuit required to prepare the quantum state is the inverse of the circuit obtained.


\begin{algorithm}
\caption{$\textsc{StatePre}(v)$}
\begin{algorithmic}[1]
\Require{A node $v$ of an LimTDD representing an $n$-qubit quantum state $\ket{\psi}$. }
\Ensure{A quantum circuit $C$, corresponding to an unitary matrix $U$, such that $U \ket{1}_a \ket{\psi}= \ket{1}_a\ket{0}^{\otimes n}$.}
\vspace{0.4em}

\State $cir \gets \mathrm{QuantumCircuit}(n+1)$ {\color{gray}\Comment{An empty quantum circuit with $n+1$ qubits}}

\If{$v$ is the terminal node}
\State \Return $(cir, 1)$
\EndIf
\State Suppose $\w\big((v,\high(v))\big)=\lambda \cdot O$
\State Append $cir$ with a controlled $O^\dag$ gate, with the control condition set to be $\ket{1}_a$  {\color{gray}\Comment{Reduce the operator on the high-edge of $v$}}

\If{$\low(v)=\high(v)$}
\State $cir_0, w_0 \gets \textsc{StatePre}\big(\low(v)\big)$
\State Append $cir$ with $cir_0$
\State $w_1 \gets w_0$
\Else
\State Use the branch condition of $\high(v)$ to close the high-branch of the node
\State $cir_0, w_0 \gets \textsc{StatePre}\big(\low(v)\big)$
\State Append $cir$ with $cir_0$
\State Use the branch condition of $v$ to close the low-branch and open the high-branch
\State $cir_1, w_1 \gets \textsc{StatePre}\big(\high(v)\big)$
\State Append $cir$ with $cir_1$
\State Use the branch condition of $\low(v)$ to open the low-branch
\EndIf
\State $w_1 \gets \lambda \cdot w_1$
\State $c \gets w_1/w_0$
\State Append a controlled $\frac{1}{\sqrt{1+|c|^2}}\left[\begin{array}{cccc} 
		1 & c^\dag\\ 
		-c & 1\\
\end{array}\right]$, with the control condition $\ket{1}_a$ and target $q_v$
\vspace{0.2em}
\State \Return $\big(cir, w_0 \cdot \sqrt{1+|c|^2}\big)$
\end{algorithmic}
\label{alg:State_Pre} 
\end{algorithm}

\section{An example}\label{sec:exp}

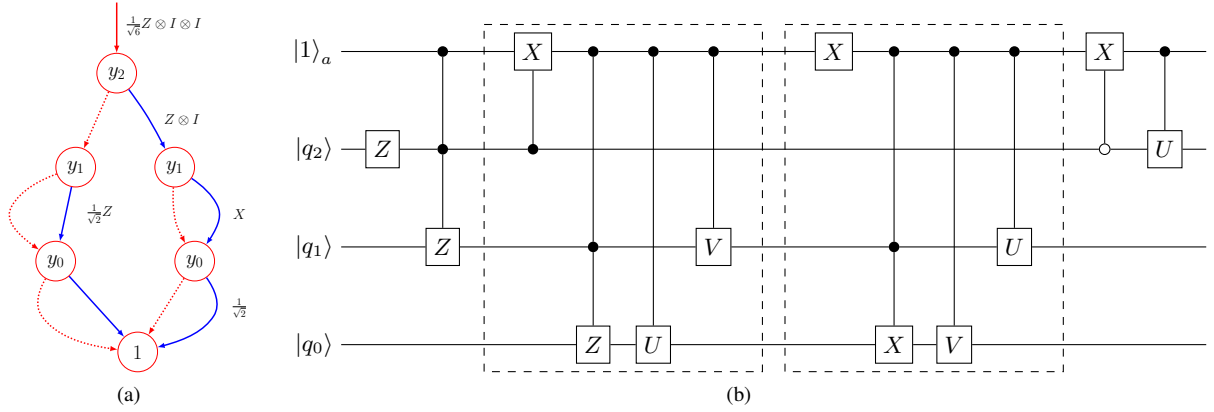
\begin{figure*}[htbp]
\centering

\subfigure[]{
    \resizebox{0.19\textwidth}{!}{




\begin{tikzpicture}[
>=latex,
line join=bevel,
every node/.style={minimum size=1.4cm, inner sep=0pt,thick, font=\LARGE}, 
every path/.style={line width=1.5pt} 
]
\node (1) at (128.67bp,18.0bp) [draw=red,circle,font=\Huge] {$1$};
  \node (y01) at (46.668bp,108.9bp) [draw=red,circle,font=\Huge] {$y_0$};
  \node (y02) at (185.67bp,108.9bp) [draw=red,circle,font=\Huge] {$y_0$};
  \node (y10) at (66.668bp,202.96bp) [draw=red,circle,font=\Huge] {$y_1$};
  \node (y11) at (165.67bp,202.96bp) [draw=red,circle,font=\Huge] {$y_1$};
  \node (y2) at (107.67bp,297.02bp) [draw=red,circle,font=\Huge] {$y_2$};
  \node (-0) at (107.67bp,387.93bp) [draw,draw=none] {};
  \draw [red,->,dotted] (y01) ..controls (29.551bp,79.314bp) and (25.128bp,64.774bp)  .. (32.668bp,54.0bp) .. controls (47.583bp,32.686bp) and (76.911bp,24.306bp)  .. (1);
  \definecolor{strokecol}{rgb}{0.0,0.0,0.0};
  \pgfsetstrokecolor{strokecol}
  \draw (53.668bp,61.875bp) node {};
  \draw [blue,->] (y01) ..controls (73.917bp,78.36bp) and (94.182bp,56.39bp)  .. (1);
  \draw (116.67bp,61.875bp) node {};
  \draw [red,->,dotted] (y02) ..controls (166.06bp,77.316bp) and (153.94bp,58.425bp)  .. (1);
  \draw (180.67bp,61.875bp) node {};
  \draw [blue,->] (y02) ..controls (205.77bp,80.723bp) and (212.16bp,65.699bp)  .. (204.67bp,54.0bp) .. controls (194.36bp,37.899bp) and (174.56bp,29.102bp)  .. (1);
  \draw (230.29bp,61.875bp) node {$\frac{1}{\sqrt{2}}$};
  \draw [red,->,dotted] (y10) ..controls (31.234bp,190.19bp) and (11.932bp,180.12bp)  .. (2.6676bp,163.81bp) .. controls (-5.0407bp,150.24bp) and (6.2358bp,136.64bp)  .. (y01);
  \draw (23.668bp,155.93bp) node {};
  \draw [blue,->] (y10) ..controls (59.707bp,169.93bp) and (56.313bp,154.3bp)  .. (y01);
  \draw (90.543bp,155.93bp) node {$\frac{1}{\sqrt{2}} Z$};
  \draw [red,->,dotted] (y11) ..controls (163.61bp,171.29bp) and (164.02bp,158.88bp)  .. (166.67bp,148.06bp) .. controls (167.5bp,144.65bp) and (168.68bp,141.19bp)  .. (y02);
  \draw (187.67bp,155.93bp) node {};
  \draw [blue,->] (y11) ..controls (193.01bp,184.7bp) and (203.51bp,175.28bp)  .. (208.67bp,163.81bp) .. controls (212.85bp,154.5bp) and (210.23bp,144.22bp)  .. (y02);
  \draw (230.42bp,155.93bp) node {$X$};
  \draw [red,->,dotted] (y2) ..controls (93.74bp,264.75bp) and (86.177bp,247.77bp)  .. (y10);
  \draw (110.67bp,249.99bp) node {};
  \draw [blue,->] (y2) ..controls (125.36bp,273.0bp) and (130.96bp,265.2bp)  .. (135.67bp,257.87bp) .. controls (140.87bp,249.76bp) and (146.13bp,240.69bp)  .. (y11);
  \draw (172.92bp,249.99bp) node {$Z\otimes I$};
  \draw [red,->] (-0) ..controls (107.67bp,358.85bp) and (107.67bp,343.46bp)  .. (y2);
  \draw (154.54bp,344.05bp) node {$\frac{1}{\sqrt{6}} Z\otimes I\otimes I$};
\end{tikzpicture}

%
%
}
}
\subfigure[]{
\begin{tikzpicture}
  \begin{yquant}[register/minimum height=12mm, operator/separation=3mm, control style={radius=2pt}, subcircuit box style={dashed}]
    qubit {$\ket{1}_a$} a;
    qubit {$\ket{q_2}$} q2;
    qubit {$\ket{q_1}$} q1;
    qubit {$\ket{q_0}$} q0;
    box {$Z$} q2;
    z q1 | q2, a;
    subcircuit {
      qubit {} a;
      qubit {} q2;
      qubit {} q1;
      qubit {} q0;
      box {$X$} a | q2;
      z q0 | q1, a;
      box {$U$} q0 | a;
      box {$V$} q1 | a;
    } (a,q2,q1,q0);
    subcircuit {
      qubit {} a;
      qubit {} q2;
      qubit {} q1;
      qubit {} q0;
      x a;
      x q0 | q1,a;
      box {$V$} q0 | a;
      box {$U$} q1 | a;
      }(a,q2,q1,q0);
    x a | ~q2;
    box {$U$} q2 | a;
  \end{yquant}
\end{tikzpicture}}
\caption{The quantum circuit that transfers the quantum state represented by the LimTDD shown in Fig. \ref{fig:limtdd} to $\ket{000}$. The two dotted boxes correspond to the processing of the $y_1$ and $y_1'$ nodes, respectively.
In this circuit $U = \frac{1}{\sqrt{2}}\left[\begin{smallmatrix} 
		1 & 1\\ 
		-1 & 1\\
\end{smallmatrix}\right]$, $V = \frac{\sqrt{2}}{\sqrt{3}}\left[\begin{smallmatrix} 
		1 & \frac{1}{\sqrt{2}}\\ 
		-\frac{1}{\sqrt{2}} & 1\\
\end{smallmatrix}\right]$.
}
\label{fig:pre_cir}
\end{figure*}

In this section, we provide a detailed example to illustrate the procedure of our algorithm. We give a step by step demonstration of our algorithm on the LimTDD given in Fig. \ref{fig:limtdd}, with the quantum state to be prepared being $ \frac{1}{\sqrt{6}} (\ket{000} + \ket{001} + \frac{1}{\sqrt{2}} \ket{010} - \frac{1}{\sqrt{2}} \ket{011} - \ket{100} - \frac{1}{\sqrt{2}} \ket{101} + \frac{1}{\sqrt{2}} \ket{110} + \ket{111}) $.

\subsection{Initial Setup}
\begin{enumerate}
    \item \textbf{Quantum State Representation}:
    The quantum state represented by the LimTDD is:
    $ \frac{1}{\sqrt{6}} Z \otimes I \otimes I \ket{y_2} $
    where:
    $ \ket{y_2} = \ket{0} \ket{y_1} + \ket{1} (Z \otimes I \ket{y'_1}) $
    $ \ket{y_1} = \ket{0} \ket{y_0} + \frac{1}{\sqrt{2}} \ket{1} (Z \ket{y_0}) $
    $ \ket{y'_1} = \ket{0} \ket{y'_0} + \ket{1} (X \ket{y'_0}) $
    $ \ket{y_0} = \ket{0} + \ket{1} $
    $ \ket{y'_0} = \ket{0} + \frac{1}{\sqrt{2}} \ket{1} $.

    \item \textbf{Ancilla Qubit}:
    We add an ancilla qubit $ \ket{1}_a $ to the system, resulting in the initial state:
    $ \frac{1}{\sqrt{6}} \ket{1}_a (Z \otimes I \otimes I \ket{y_2}) $.
\end{enumerate}

\subsection{Step-by-Step Reduction}
\begin{enumerate}
    \item \textbf{Cancel the Operator on the Incoming Edge}:
    \begin{itemize}
        \item Apply a $Z$ gate on qubit $ q_2 $ to cancel the $ Z \otimes I \otimes I $ operator on the incoming edge of the LimTDD. The state becomes:
        $ \ket{1}_a \ket{y_2} $.
    \end{itemize}

    \item \textbf{Process the High-Edge of $ y_2 $ Node}:
    \begin{itemize}
        \item Apply a CCZ gate with $ q_a $ and $ q_2 $ as controls and $ q_1 $ as the target to cancel the $ Z \otimes I $ operator on the high-edge of the $ y_2 $ node. The state becomes:
    $ \ket{1}_a (\ket{0} \ket{y_1} + \ket{1} \ket{y'_1}) $.
    \end{itemize}

    \item \textbf{Process the $ y_1 $ Node}:
    \begin{itemize}
        \item Use a $CX$ gate with $ q_2 $ as the control qubit and $ q_a $ as the target qubit to close the $ y'_1 $ node. The state becomes:
        $ \ket{1}_a \ket{0} \ket{y_1} + \ket{0}_a \ket{1} \ket{y'_1} $.
        \item Apply a CCZ gate with $ q_a $ and $ q_1 $ as controls and $ q_0 $ as the target to cancel the $ Z$ operator on the high-edge of the $ y_1 $ node. The state becomes:
        $ \ket{1}_a \ket{0} (\ket{0} + \frac{1}{\sqrt{2}} \ket{1}) \ket{y_0} + \ket{0}_a\ket{1} \ket{y'_1} $.
    \end{itemize}

    \item \textbf{Process the $ y_0 $ Node and Adjust the Weights on Outgoing Edges of the $y_1$ Node}:
    \begin{itemize}
        \item Since $ \ket{y_0} = \ket{0} + \ket{1} $, apply a controlled-U gate with $ q_a $ as the control qubit to transform the state to:
        $ \ket{1}_a \ket{0} (\sqrt{2}\ket{0} + \ket{1}) \ket{0} + \ket{0}_a \ket{1} \ket{y'_1}  $
        where $ U = \frac{1}{\sqrt{2}} \begin{bmatrix} 1 & 1 \\ -1 & 1 \end{bmatrix} $.
        \item Apply a controlled-$V$ gate with $ q_a $ as the control qubit to adjust the weights on the outgoing edges of the $ y_1 $ node and change the state to: $\sqrt{3}\ket{1}_a\ket{0}\ket{0}\ket{0}+\ket{0}_a\ket{1}\ket{y_1^{\prime}}$
        where $ V = \frac{\sqrt{2}}{\sqrt{3}}\left[\begin{array}{cccc} 
		1 & \frac{1}{\sqrt{2}}\\ 
		-\frac{1}{\sqrt{2}} & 1\\
\end{array}\right]$.
    \end{itemize}


    \item \textbf{Process the $ y'_1 $ Node}:
    \begin{itemize}
        \item Use an $X$ gate on $ q_a $ to switch the branches and open the $ y'_1 $ node. The state becomes:
        $ \sqrt{3} \ket{0}_a \ket{0} \ket{0} \ket{0}  + \ket{1}_a \ket{1} \ket{y'_1} $.
        \item Apply a CCX gate controlled by $ q_a $ and $ q_1 $ to cancel the operator on the high-edge of the $ y'_1 $ node. The state becomes:
        $ \sqrt{3} \ket{0}_a \ket{0} \ket{0} \ket{0}  + \sqrt{3} \ket{1}_a \ket{1} (\ket{0}+\ket{1}) \ket{y_0'} $.
    \end{itemize}

    \item \textbf{Process the $ y_0' $ Node and Adjust the Weights on Outgoing Edges of the $y_1'$ Node}:
    \begin{itemize}
        \item Apply a controlled-$V$ gate with $ q_a $ as the control qubit, the state will be changed to 
$\sqrt{3}\ket{0}_a\ket{0}\ket{0}\ket{0}+\ket{1}_a\ket{1}\big(\frac{\sqrt{3}}{\sqrt{2}}\ket{0}+\frac{\sqrt{3}}{\sqrt{2}}\ket{1}\big)\ket{0}$.
        \item Apply a further gate $U$ controlled by $\ket{1}_a$ and change the state to 
$\sqrt{3}\ket{0}_a\ket{0}\ket{0}\ket{0}+\sqrt{3}\ket{1}_a\ket{1}\ket{0}\ket{0}$.
    \end{itemize}

    \item \textbf{Adjust the Weights on Outgoing Edges of the $y_2$ Node}:
    \begin{itemize}
        \item Apply a CX gate with the control qubit $q_2$ set to be $\ket{0}$ and target qubit $q_a$, to open the $y_1$ node, thus making the all branches of $y_2$ node open, and the state becomes: $\sqrt{3}\ket{1}_a(\ket{0}+\ket{1})\ket{0}\ket{0}$.
        \item Use a controlled-$U$ gate with $ q_a $ as the control qubit to adjust the weights on the outgoing edges of the $ y_2 $ node, the state becomes: 
        $ \sqrt{6} \ket{1}_a\ket{0} \ket{0} \ket{0} $. Note that the coefficient $\sqrt{6}$ is cancelled with the ignored coefficient $\frac{1}{\sqrt{6}}$.
    \end{itemize}
\end{enumerate}

\subsection{Resulting Quantum Circuit}
The resulting quantum circuit for preparing the desired quantum state is shown in Fig. \ref{fig:pre_cir}, which includes the following gates:
\begin{itemize}
    \item $Z$ gates to cancel the operators on the incoming edges.
    \item CCZ and CCX gates to cancel the operators on the high-edges of the nodes.
    \item Controlled-$U$ and controlled-$V$ gates to adjust the weights on the edges.
    \item Ancilla qubit $ q_a $ used to control the processing of different branches.
\end{itemize}

\section{Complexity}\label{sec:complexity}

\subsection{Time complexity}

In our algorithm, we first process the operator on the incoming edge of the root node of the LimTDD, then traverse the decision diagram starting from the root node and subsequently handle the operator on the high-edge of each node. When we encounter a branch node (that is, a node whose 0- and 1-successors are different), we use an auxiliary qubit to mark and close its 1-successor, then proceed to process the 0-successor. After completing the processing of the 0-successor, we return to the branch node, close the 0-successor, and open the 1-successor for processing. If the current node is not a branch node, we simply process its successors without closing half of the decision diagram. This process continues recursively.

Upon reaching the terminal node, we perform a reverse pass, processing the complex weights on each node's edges in a bottom-up manner. After reaching the nearest branch node, we wait for the processing of branch 1 to complete, then process the complex weight on the edge of the branch node, and subsequently backtrack upwards. Eventually, all operators and complex weights on the branches will have been resolved.


To summarise, we traverse the decision graph in a depth-first manner, with each reduced path being traversed exactly once. Note that there are $n$ non-terminal nodes on a path (where $n$ is the number of qubits in the system), each requiring handling of the high-edge operator and outgoing weight. Assuming there are $p$ reduced paths in the LimTDD, the time complexity of our algorithm is $\mathcal{O}(np)$.

\subsection{Gate Complexity}

We begin by analysing the operators appearing on the edges of the LimTDD. There are no more than $(n-1)+\cdots+2+1 = \frac{n(n-1)}{2} $ local operators on each path, and these operators will be eliminated through 3-qubit controlled gates. Thus, we need $\mathcal{O}(n^2p)$ 3-qubit gates to eliminate all these operators. Note that since the control qubits to cancel the local operators on an edge are all the same, these 3-qubit gates can be reduced to two 3-qubit gates and a series of 2-qubit gates if a further ancilla qubit is given. In addition, we need no more than $n$ single-qubit gates to eliminate the operator on the incoming edge of the LimTDD.

Next, we examine the weights on the outgoing edges of each node. As each node on each path requires an operator, and each operator requires a control qubit, we need $\mathcal{O}(np)$ 2-qubit gates to eliminate them.

Finally, we examine the quantum gates used to control the opening and closing of branches. For each branch node, we need one controlled gate to close its 1-branch, one controlled gate to flip the open/close status of both branches, and one controlled gate to reopen the 0-branch. Assuming there are $k$ branch nodes preceding the current node on the path, this requires two $(k+1)$-qubit controlled gates and one $k$-qubit controlled gate. 


In summary, the total gate requirements are:
\begin{itemize}
    \item $\mathcal{O}(p)$ $s$-qubit gates for $s>3$,
    \item $\mathcal{O}(n^2p)$ 3-qubit gates,
    \item $\mathcal{O}(np)$ 2-qubit gates,
    \item $\mathcal{O}(n)$ single-qubit gates.
\end{itemize}

\textbf{Special Case:} For decision diagrams in the tower form (where the 0- and 1-successors of each non-terminal node are the same), there is only one path involving different nodes. In this scenario, the ancilla qubit can be omitted since it remains in the $\ket{1}$ state. Consequently, only $\mathcal{O}(n^2)$ 2-qubit gates and $\mathcal{O}(n)$ single-qubit gates are required to prepare the quantum state. 

\section{Experiments}\label{sec:experiments}

In this section, we evaluate the performance of our LimTDD-based QSP algorithm and compare it with existing methods, including the ADD-based method \cite{mozafari_efficient_2022}, Qiskit \cite{qiskit2024}, and QuICT \cite{quict}. The algorithms used for quantum state preparation in Qiskit and QuICT are established in \cite{Iten_Colbeck_Kukuljan_Home_Christandl_2016} and \cite{mottonen2004transformation}, respectively. The experiments focus on two main metrics: gate complexity and runtime complexity.

\subsection{Experimental Setup}

\begin{itemize}
    \item \textbf{Hardware and Software}:
    \begin{itemize}
        \item Our Method, Qiskit, and QuICT: Experiments are conducted on a desktop with an 11th Gen Intel Core(TM) i7-11700F CPU and 16GB RAM. All methods are implemented using the Python programming language.
        
        \item ADD-based method: Since the implementation of \cite{mozafari_efficient_2022} is in C++ and only supports a Linux environment, these experiments are conducted on a Linux server with a 13th Gen Intel(R) Core(TM) i5-13600KF and 32GB RAM.
    \end{itemize}
    \item \textbf{Quantum States}:
    \begin{itemize}
        \item We generate random quantum states using Clifford + T circuits, which are one of the most commonly used circuit categories. These states serve as the target states for the QSP task.
        \item For each qubit number $n$, we generate 20 random quantum states and calculate the average performance metrics.
    \end{itemize}
    \item \textbf{Metrics}:
    \begin{itemize}
        \item Gate Complexity: The number of multi-qubit gates required to prepare the target state. The output circuit of Qiskit and QuICT consists of single-qubit gates and $CX$ gate. For our method, we also compile the generated circuit into single-qubit gates and $CX$ gate using Qiskit and count its $CX$ gate count (denoted as ``Transpiled''). For the ADD-based method, we used the method given in \cite{mozafari_efficient_2022} (integrated in their tool) to estimate the number of $CX$ gates that needed to implement the multi-qubit gates (denoted as ``Expanded'').
        \item Runtime Complexity: The execution time required to generate the quantum circuit for state preparation. The runtime of LimTDD (Transpiled) also includes the transpilation time.
    \end{itemize}    
\end{itemize}



\subsection{Comparison with ADD-based Method}

\begin{figure*}
    \centering
    \includegraphics[width=0.4\linewidth]{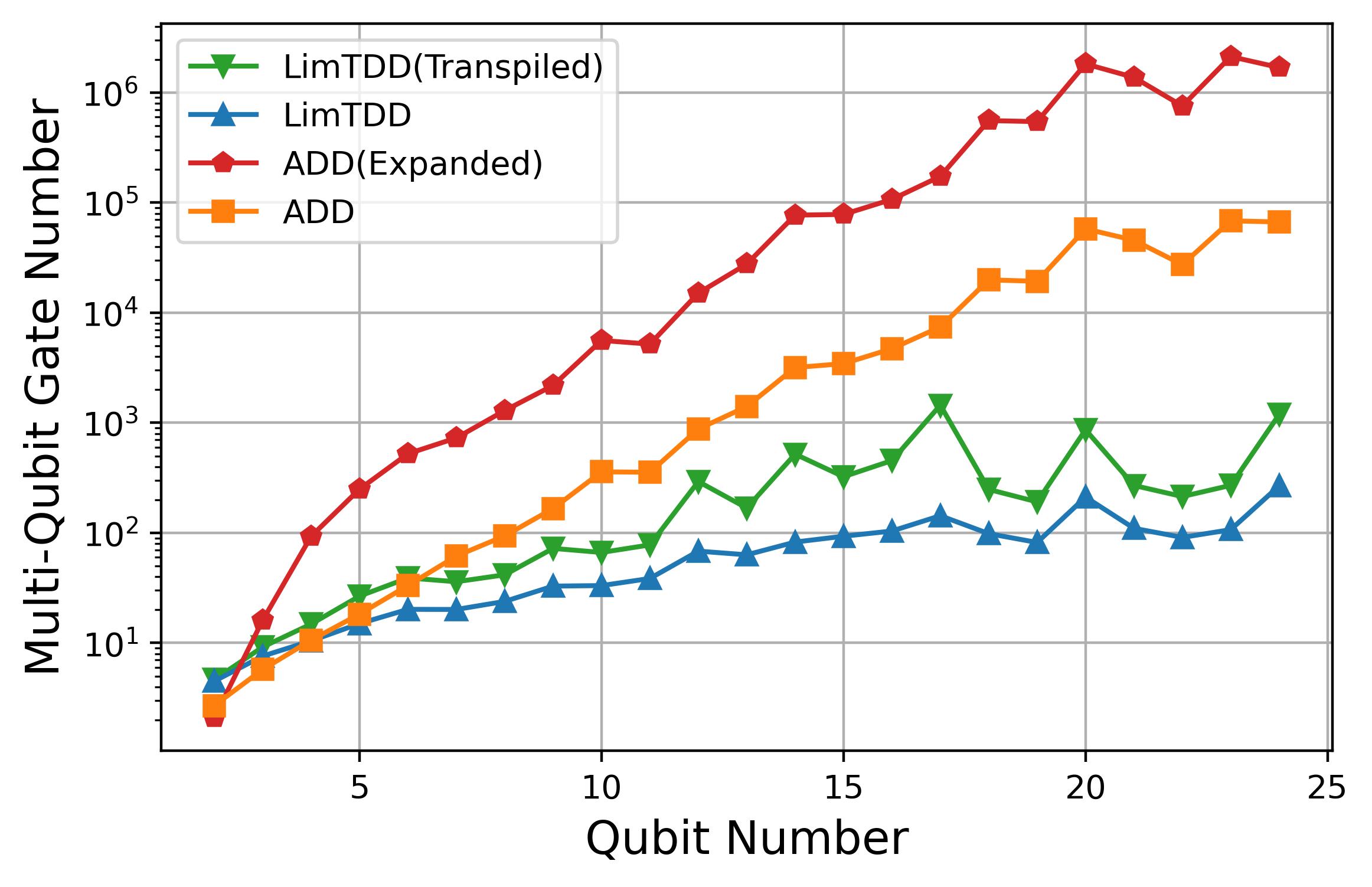}
    \includegraphics[width=0.4\linewidth]{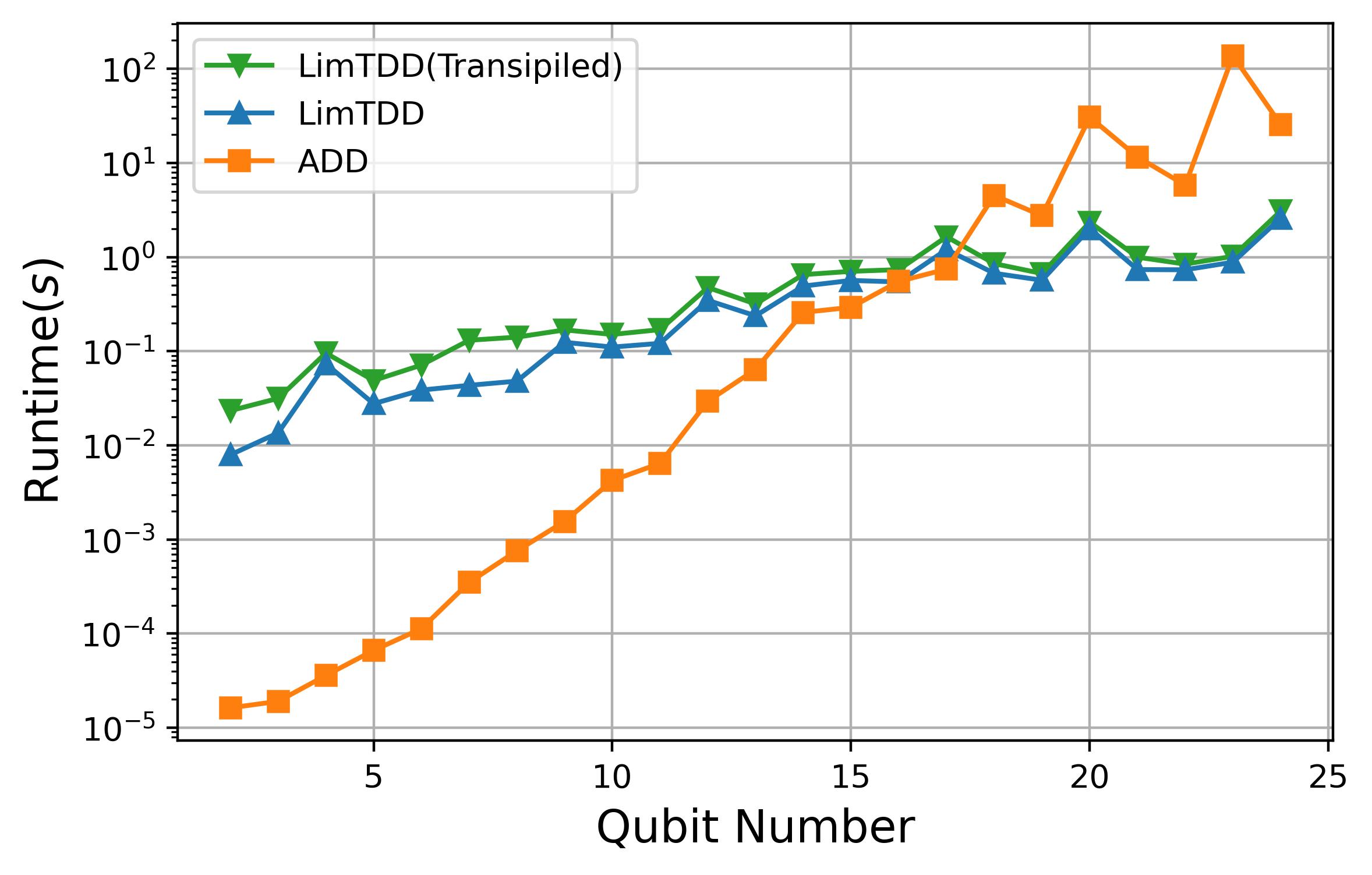}    
    \caption{Experiment results of our method against ADD-based method \cite{mozafari_efficient_2022}.}
    \label{fig:vs_ADD}
\end{figure*}

We first compare our LimTDD-based method with the ADD-based method proposed in \cite{mozafari_efficient_2022}, which also focuses on decision-diagram-based efficient state preparation. The runtime and the multi-qubit gates complexity of the two methods are demonstrated in Fig. \ref{fig:vs_ADD}.


\begin{itemize}
    \item \textbf{Gate Complexity}:
     We compare the number of gates required by our LimTDD-based method and the ADD-based method from \cite{mozafari_efficient_2022}. The results show that our method consistently requires fewer gates when the number of qubits is larger than 5. This is due to the more compact representation of quantum states using LimTDD compared to ADD. For example, for $n = 15$ qubits, our method uses approximately 100 and 700 gates before and after the transpilation, while the ADD-based method uses around 3500 and 80000 gates before and after the expansion.

    \item \textbf{Runtime Complexity}:
     When the number of qubits is small, the ADD-based method is faster than our method. The main reason could be that the ADD-based method is implemented in C++, while our method is implemented in Python. But when the number of qubits grows larger, the runtime of our method becomes significantly shorter. At this point, the advantage of the compactness of LimTDD begins to manifest.
\end{itemize}

\subsection{Comparison with Qiskit and QuICT }

\begin{figure*}
    \centering  
    \includegraphics[width=0.4\linewidth]{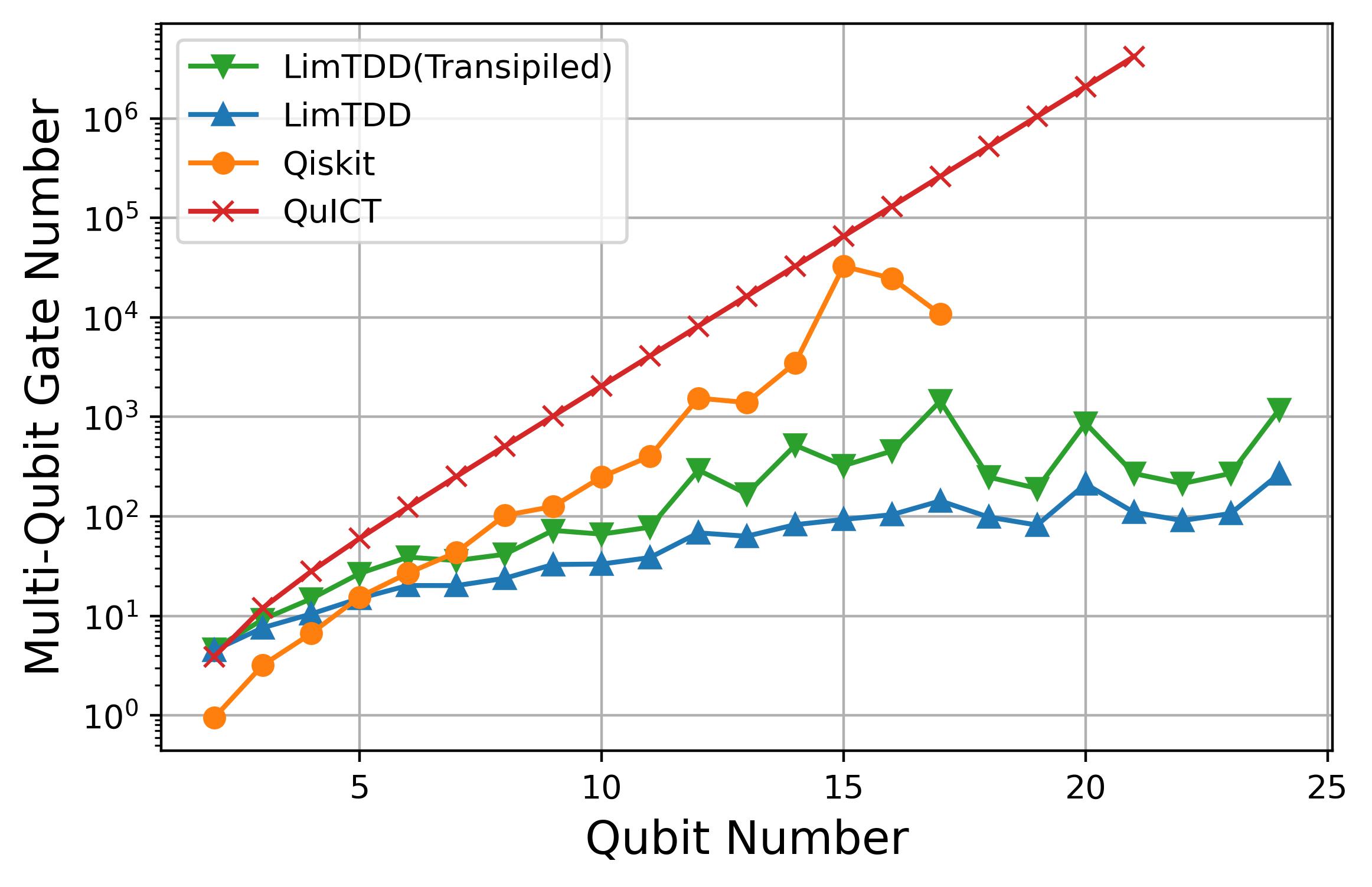}
    \includegraphics[width=0.4\linewidth]{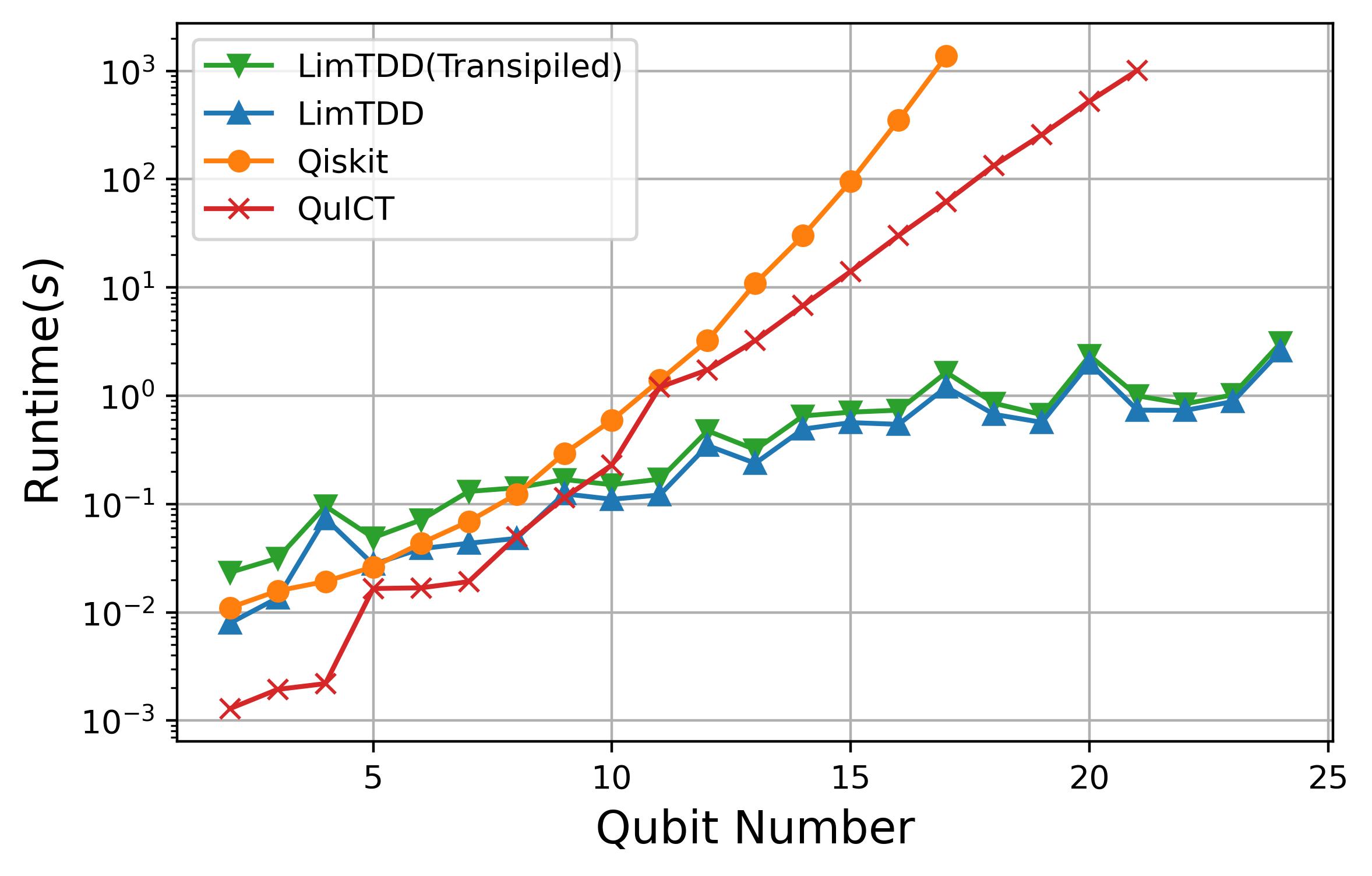}
    \caption{Experiment results of our method against Qiskit \cite{qiskit2024} and QuICT \cite{quict}.}
    \label{fig:vs_Qis_ICT}
\end{figure*}

We further compare our LimTDD-based method with two widely-used quantum computing frameworks: Qiskit \cite{qiskit2024} and QuICT \cite{quict}. Fig. \ref{fig:vs_Qis_ICT} illustrates the detailed comparison. The overall result is consistent with that of ADD-based method. We note here that no ancilla qubits are used by Qiskit and QuICT.


\begin{itemize}
    \item \textbf{Gate Complexity}:
We compare our method with Qiskit and QuICT in terms of the number of multi-qubit gates required. Our method consistently achieves lower gate counts when the number of qubits is larger than 7, demonstrating its efficiency in QSP. For instance, for $n=15$ qubits, our method requires around 100 and 700 gates before and after the transpilation, while both Qiskit and QuICT require around 130000 gates.

    \item \textbf{Runtime Complexity}:
The runtime of our method is significantly shorter compared to Qiskit and QuICT, when the number of qubits grows larger than 10. For $n=12$ qubits, our method completes in an average of 0.5 seconds, while Qiskit and QuICT take approximately 3 and 2 seconds, respectively. In addition, Qiskit and QuICT run out of time ($>600$ seconds) when the qubit number is larger than 17 and 21, respectively.

\end{itemize}

\subsection{Discussion}




To highlight the exponential advantage of our method, we provide a simple example that demonstrates the superiority of our LimTDD-based algorithm over the ADD-based method presented in \cite{mozafari_efficient_2022}.

Consider a quantum circuit extended from the one shown in Fig. \ref{fig:quantum_circuit}. Suppose $n$ is the number of qubits, starting from the state $\ket{0}^{\otimes n}$. Apply a Hadamard gate ($H$) to each qubit, and then apply a series of controlled-$Z$ (CZ) gates between qubit $q_k$ and qubits $q_0, \dots, q_{k-1}$ for $1\leq k\leq n-1$. Denote the resulting quantum state as $\ket{v_n}$. Then,
\[
\ket{v_n} = \frac{1}{\sqrt{2}} \ket{0} \ket{v_{n-1}} + \frac{1}{\sqrt{2}} \ket{1} (Z \otimes \cdots \otimes Z \ket{v_{n-1}}).
\]
For all $k = n$ down to $k = 2$, the first part (corresponding to $\ket{0}$) and the second part (corresponding to $\ket{1}$) of the state are different. Consequently, the ADD representation of the state has an exponential number of paths, while the LimTDD representation has only $n+1$ nodes, with all the operators $Z \otimes \cdots \otimes Z$ appearing on the high edge of every non-terminal node, and one path.

Using our algorithm to generate a circuit for preparing the state $\ket{v_n}$ results in a similar circuit as shown in Fig. \ref{fig:quantum_circuit}, requiring $\mathcal{O}(n^2)$ gates. An ancilla qubit $q_a$ is added, and all gates become controlled versions with $q_a$ as the control qubit. Since the state of $q_a$ remains in $\ket{1}$ and never changes, a simple optimisation can restore the circuit. In contrast, the exponential number of paths in the ADD representation necessitates an exponential number of gates for state preparation using the method of \cite{mozafari_efficient_2022}. In fact, an exponential number of quantum gates is also needed for Qiskit and QuICT to prepare this quantum state.

This example underscores the potential of LimTDD to handle complex quantum states more efficiently, providing a significant advantage in certain scenarios.

\section{Conclusion}\label{sec:conclusion}

This paper presents a novel quantum state preparation algorithm using LimTDD, which achieves significant improvements in efficiency and circuit complexity compared to existing methods. The compactness of LimTDD allows for efficient handling of large-scale quantum states, with experimental results demonstrating exponential efficiency gains in the best-case scenario. This work highlights the potential of LimTDD in quantum state preparation and provides a robust foundation for future quantum computing technologies. Future work may focus on further optimising LimTDD and exploring its applications in other quantum computing areas.







\bibliographystyle{IEEEtran}
\bibliography{reference}

\end{document}